\documentclass[notitlepage,showkeys,twocolumn,11pt,times,superscriptaddress,a4paper]{revtex4-2}
\usepackage{amsmath,amsfonts,mathtools}
\usepackage{stix}
\usepackage{physics}
\usepackage[english]{babel}
\usepackage{graphicx}
\usepackage{dcolumn}
\usepackage{bm}
\usepackage{xcolor}
\usepackage{geometry}
\usepackage[colorlinks=true,
            linkcolor=blue,
            urlcolor=blue,
            citecolor=blue]{hyperref}

\begin{document}


\title{Characterizing and quantifying weak chaos in fractional dynamics} 


\author{Daniel Borin}
\email[]{d.borin@hotmail.com}
\affiliation{University of North Dakota, School of Electrical Engineering and Computer Science, 58202, Grand Forks, ND, USA}
\affiliation{São Paulo State University (UNESP), Institute of
Geosciences and Exact Sciences, 13506-900, Rio Claro, SP, Brazil}

\author{José Danilo Szezech Jr.}
\affiliation{Graduate Program in Science, State University of Ponta Grossa, 84030-900, Ponta Grossa, PR, Brazil}
\affiliation{Department of Mathematics and Statistics, State University of Ponta Grossa, 84030-900, Ponta Grossa, PR, Brazil}
\affiliation{Institute of Physics, University of São Paulo, 05508-900, São Paulo, SP, Brazil}

\author{Matheus Rolim Sales}
\affiliation{São Paulo State University (UNESP), Institute of
Geosciences and Exact Sciences, 13506-900, Rio Claro, SP, Brazil}
\affiliation{University of Essex, School of Mathematics, Statistics and Actuarial Science, Wivenhoe Park, Colchester, CO4 3SQ, UK}



\begin{abstract}
A particularly intriguing and unique feature of fractional dynamical systems is the cascade of bifurcations type trajectories (CBTT). We examine the CBTTs in a generalized version of the standard map that incorporates the Riemann-Liouville fractional derivative, known as the Riemann-Liouville Fractional Standard Map (RLFSM). 
We propose a methodology that uses two quantifiers based solely on the system's time series: the Hurst exponent and the recurrence time entropy, for characterizing such dynamics. This approach allows us to effectively characterize the dynamics of the RLFSM, including regions of CBTT and chaotic behavior. Our analysis demonstrates that regions of CBTT are associated with trajectories that exhibit lower values of these quantifiers compared to strong chaotic regions, indicating weakly chaotic dynamics during the CBTTs.
\end{abstract}


\maketitle


\section{Introduction}
Fractional dynamical systems (FDS) are systems governed by fractional differential/difference equations (FDE), incorporating fractional time derivatives/differences \cite{MACHADO20111140,letnikov1868historical,valerio2014some}. FDEs are integro-differential/difference equations, and solving them requires extensive computational resources \cite{Edelman2014on}. 
This complexity makes the investigation of their general properties particularly challenging. In this sense, the research on nonlinear fractional dynamical systems has advanced notably, with several studies \cite{edelman2023stability, edelman2022asymptotic, khennaoui2023lozi, ding2023controllable, MA2023114113, doi:10.1142/S0218127423501973} [38–44]. Many of these investigations are based on fractional maps that describe periodically kicked systems. \cite{doi:10.1142/S0218127424500858, fractalfract9070472, e20100720, 10.1063/1.3443235,  Tarasov2008, Tarasov2009,10.1063/1.3272791}.


A special property of these systems is that FDS exhibit memory effects, leading to potentially unconventional properties in FDE solutions \cite{Edelman2020,10.1063/5.0151812, 10.1063/1.4819165}: trajectories may intersect, attractors can overlap, and attractors exist only in an asymptotic sense, with their limiting values not necessarily belonging to their basins of attraction. 

In this sense, a novel type of regime unique to FDS emerges in fractional systems, called cascade of bifurcations type trajectories (CBTT) \cite{10.1063/1.4819165}. In CBTT, a sequence of bifurcations occurs not due to variations in the system parameters, as in traditional dynamical systems, but rather along a single attracting trajectory during its temporal evolution. These bifurcations occur within specific time windows during which the trajectory is confined to smaller subsets of the phase space, as if temporarily trapped. The orbit enters from the chaotic region into a CBBT, remains there for some time, and then returns to the chaotic region, repeating this process multiple times. This behavior is similar to that of sticky orbits in two-dimensional, area-preserving maps. The stickiness effect \cite{contopoulos1971orbits, AfraZas98, PhysRevResearch.5.043196, MEISS1983375, ZASLAVSKY2002461, 10.1063/1.1979211, PhysRevE.73.026207, PhysRevLett.100.184101}, also known as weak chaos, is one of the main features of two-dimensional, area-preserving maps. It occurs when chaotic orbits exhibit prolonged interactions with specific regions in phase space (stability islands) that temporarily ``trap'' these orbits, making them almost like to quasiperiodic orbits. These trappings, however, do not make a chaotic orbit a regular one. The orbit still exhibits a positive largest Lyapunov exponent but smaller in comparison when not trapped \cite{SZEZECH2005394}.

Various methods have been proposed to quantify the stickiness effect, including finite-time Lyapunov exponents \cite{Okushima2003_PRL, SZEZECH2005394, PhysRevE.99.052208, PhysRevE.91.062907, HARLE2007130}, the distribution of Poincaré recurrence times \cite{ARTUSO199968,Altmann_2007}, measures based on recurrence quantification analysis \cite{10.1063/1.2785159, Palmero10.1063/5.0102424}, weighted Birkhoff averages \cite{SALES2022127991,SANDER2020132569}, and finite-time rotation number \cite{SANTOS10.1063/1.5078533}. Recently, two methods have emerged for rapidly detecting these dynamical traps: the Shannon entropy of recurrence times (recurrence time entropy) \cite{SALES10.1063/5.0140613, PhysRevE.109.015202} and the Hurst exponent \cite{Borin_Hurst}.

In this paper, we introduce a methodology that uses time series analysis of the Hurst exponent and Recurrence Time Entropy (RTE) to provide a more refined characterization of CBTT in fractional systems. We apply this approach to a generalization of the standard map that incorporates the fractional derivative of Riemann-Liouville into the equations of motion. Our results show that regions of CBTT are associated with trajectories that exhibit lower values of these quantifiers. This method is general and can also be applied to Hamiltonian systems with ordinary derivatives.

The paper is organized as follows: Section \ref{sec:met} provides a brief overview of the Hurst Exponent and Recurrence Time Entropy (RTE), including the algorithms used for their calculation. Section \ref{sec:mod} describes the Riemann-Liouville Fractional Standard Map (RLFSM) and discusses some properties of fractional maps. Section \ref{sec:results} presents the main results, illustrating how the proposed methodology enables the Hurst Exponent and RTE to effectively identify weak chaos regions in fractional maps characterized by CBTT. Finally, Section \ref{sec:conl} offers a summary of the main findings and final remarks.

\section{Methods} \label{sec:met}

In this section, we outline the methodologies employed to calculate the measures used in this paper to quantify the CBTTs: the Hurst Exponent and the recurrence time entropy (RTE). 

\subsection{Hurst exponent}\label{sec:HE}

The Hurst exponent, introduced by H. E. Hurst in 1951 to statistically model the cyclical patterns of Nile floods \cite{Hurst51}, serve as a fundamental measure of long-term memory in time series. Its applicability spans various domains, including financial market analysis \cite{MATOS20083910,COUILLARD2005404,CARBONE2004267,GRECH2004133}, electrocardiogram data classification for heart disease \cite{LAHMIRI2023100142,sridhar2021accurate}, climate temperature \cite{10.1063/5.0206846}, and even experimental measurement in specific contexts \cite{PhysRevLett.94.010602}.

A wide array of computational algorithms are available \cite{zhang2023surveymethodsestimatinghurst} for estimating the Hurst exponent, such as detrended fluctuation analysis (DFA) \cite{PhysRevE.49.1685}, detrended moving average (DMA) \cite{DMA}, and periodogram method (PM) \cite{PM}, to cite a few. Among them, the rescaled range analysis (R/S analysis) stands out as the oldest and most renowned method \cite{Hurst51}, popularized by Mandelbrot and Wallis' works \cite{MW68, MW69}. 

In this approach, given a time series $\vec{x} = (x_1, x_2, \ldots, x_{N})$ of length $N$, then:

\begin{enumerate}
	\item Divide the time series into \( \kappa \) subseries \( P_{k,\ell} \) of length \( \ell \), such that the number of chunks $\kappa$ satisfies $\kappa=N/\ell$. Each subseries is denoted by $P_{k,\ell}=[x_{(k-1)\ell+1}, x_{{k \ell}}]$ with $k =1,2,\hdots, \kappa$.
	\item For each subseries \( k = 1, 2, \ldots, \kappa \), calculate the mean \( \mu_{k,\ell} \), standard deviation $S_{k,\ell}$ and the deviations from the mean:
	\[
	D_{i,k,\ell} = P_{i,k,\ell} - \mu_{k,\ell}
	\]
	where $i$ denotes the elements.
	
	\item Compute cumulative sums of deviations:
	\[
	Z_{i,k,\ell} = \sum_{j=1}^{i} D_{j,k,\ell}
	\]
	for \( i = 1, 2, \ldots, \ell \).
	
	\item Calculate the range of the cumulative deviation \( R_{k,\ell} \) of each subseries \( Z_{i,k,\ell} \).
	
	\[
	R_{k,\ell} = \max\limits_{1<i<\ell} \left(Z_{i,k,\ell}\right)-\min\limits_{1<i<\ell} \left(Z_{i,k,\ell}\right),
	\]

	\item Calculate the mean of the rescaled ranges:

	\[(R/S)_\ell = \left\langle\dfrac{R_{k,\ell}}{S_{k,\ell}}\right\rangle_k = \dfrac{1}{\kappa} \sum_{k=1}^{\kappa}\dfrac{R_{k,\ell}}{S_{k,\ell}}\]

	\item Repeat the process considering another value for $\ell$, that is, dividing the time series into another number of subseries.
	
	\item Estimate the Hurst exponent \( H \) by assuming a power-law relationship:
	\[
	(R/S)_\ell =C \ell^H
	\]
	and using regression analysis to find \( H \).
\end{enumerate}

\subsection{Recurrence time entropy}
\label{subsec:recplot}

\begin{figure}[h]
	\includegraphics[width=\linewidth]{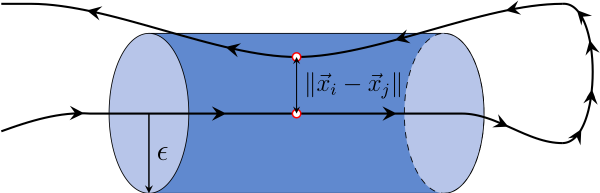}
	\caption{A diagram illustrating the recurrence analysis, in which each point in the recurrence plot (RP) corresponds to a segment of the trajectory that remains within an $\epsilon$-neighborhood of another segment.}
	\label{fig:recurrence_scheme}
\end{figure}

The recurrence plot (RP), introduced by Eckmann \textit{et al.} in 1987 \cite{Eckmann_1987}, is a graphical tool used to visualize the recurrences of a time series in the $d$-dimensional phase space of a dynamical system. For a trajectory $\vec{x}_i \in \mathbb{R}^d$ ($i = 1, 2, \ldots, N)$ of length $N$, the $N \times N$ recurrence matrix is defined as
\begin{equation}
	\label{eq:recmat}
	R_{ij} = H\left( \epsilon - \|\vec{x}_i - \vec{x}_j\| \right),
\end{equation}
where $i, j = 1, 2, \ldots, N$, $H(\cdot)$ is the Heaviside unit step function, $\epsilon$ is a threshold, and $\|\vec{x}_i - \vec{x}_j\|$ denotes the spatial distance between two states, $\vec{x}_i$ and $\vec{x}_j$, in phase space, measured using an appropriate norm, which in this work is taken to be the $L_\infty$-norm (maximum norm). Although the $L_2$-norm (Euclidean norm) yields similar RP \cite{MARWAN2007237}, for a fixed threshold $\epsilon$ the maximum norm finds the most recurrent points and it is computationally faster. Hence we prefer the maximum norm. Essentially, recurrence refers to a trajectory returning near a previously visited state, as illustrated in Fig. \ref{fig:recurrence_scheme}.

\begin{figure*}[t]
	\includegraphics[width=\linewidth]{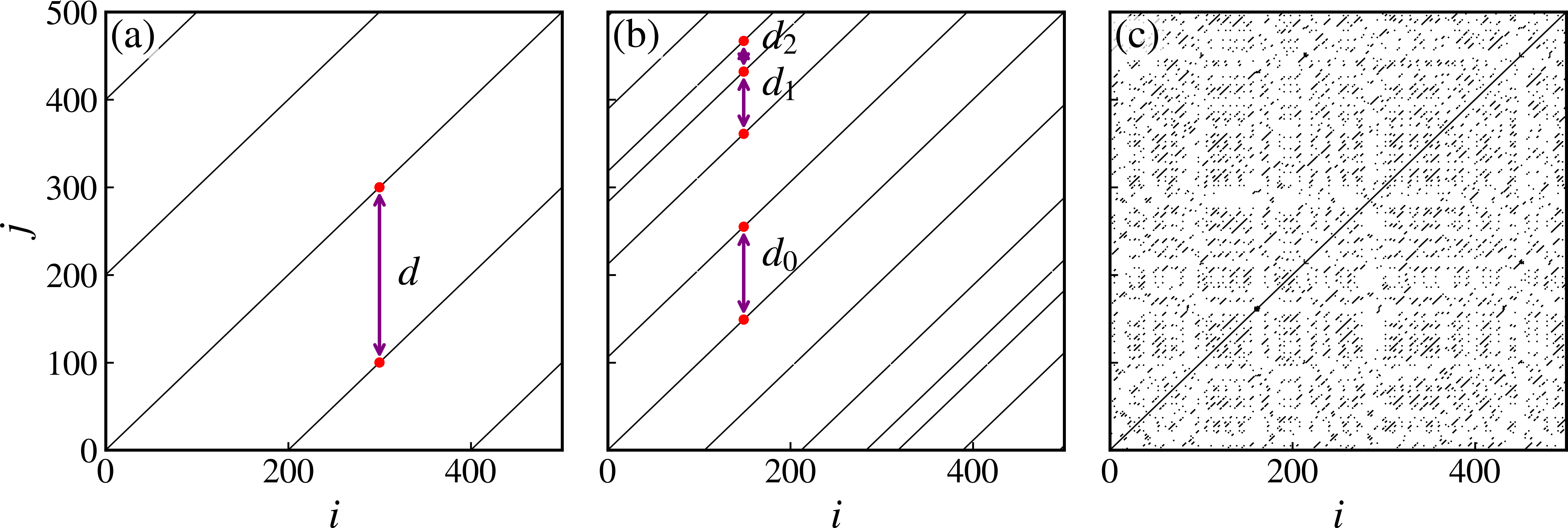}
	\caption{RPs of four three cases: (a) periodic, (b) quasiperiodic, and (c) chaotic. The purple line with double arrows in (a) and (b) denote the white vertical lines.}
	\label{fig:recurrence_plots}
\end{figure*}

The recurrence matrix \(\mathbf{R}\) is a symmetric, binary matrix where recurrent states are represented by a value of 1 and non-recurrent states by a value of 0. Two states are considered recurrent if they are ``close'' to each other within a distance \(\epsilon\), meaning \(\vec{x}_i \approx \vec{x}_j\). Therefore, the choice of \(\epsilon\) is crucial and not arbitrary. If \(\epsilon\) is too large, nearly every point will be recurrent with every other point. Conversely, if \(\epsilon\) is too small, there will be almost no recurrent states. Several methods for selecting \(\epsilon\) have been proposed. Some methods fix \(\epsilon\) based on a desired recurrence point density in the RP \cite{ZBILUT2002173}, while others use \(\epsilon\) as a fraction of the standard deviation \(\sigma\) of the time series \cite{MARWAN2007237, THIEL2002138, Schinkel2008}. In this study, we set the threshold to \(10\%\) of the time series standard deviation, which has been proven effective for detecting stickiness in two-dimensional area-preserving maps \cite{SALES10.1063/5.0140613,PhysRevE.109.015202}. 

Graphically, RP is a visualization of the recurrence matrix, where each recurrent state (a pair $(i, j)$ such that $R_{ij} = 1$) is shown as a colored dot. This visual representation reveals different patterns depending on the trajectory's evolution, which is determined by its initial conditions. In Fig.\ref{fig:recurrence_plots}, we present examples of distinct RP patterns. Periodic motion produces long, uninterrupted diagonal lines, regularly spaced by a distance $d$, as shown in Fig.\ref{fig:recurrence_plots}(a). The vertical distance between these lines corresponds to the recurrence time, meaning that for periodic dynamics, $d$ represents the oscillation period. Quasiperiodic motion, illustrated in Fig.~\ref{fig:recurrence_plots}(b), also produces mainly uninterrupted diagonal lines, but unlike the periodic case, the spacing between the diagonals varies (denoted by $d_0$, $d_1$ and $d_2$), indicating the presence of multiple return times. Finally, Fig.~\ref{fig:recurrence_plots}(c) shows the RP of a chaotic trajectory, which exhibits disrupted and irregular line structures. The varying distances between diagonals reflect the multiple time scales in the system, while the broken lines result from the exponential divergence of nearby trajectories \cite{Zou2007a}.


Several measures have been proposed to characterize and quantify the structures in RPs. Some of them include the recurrence rate, the determinism, and the laminarity, to cite a few. We refer the reader to Refs. \cite{MARWAN2007237, PhysRevE.66.026702, MARWAN2002299,Marwan2008} for a complete discussion on these and other measures. Entropy-based measures have also been employed to quantify RPs that allow the identification of chaotic regimes and bifurcation points \cite{WZ94,Baptista2010,TRULLA1996255,PhysRevE.90.042919}. One particular entropy-based measure relies on the estimation of the recurrence times of a trajectory using its corresponding RP. The vertical distance between the diagonal lines (white vertical lines), \textit{i.e.}, the gaps between them, are an estimate of the trajectory recurrence times \cite{Zou2007a, Zou2007b, Baptista2010,Ngamga2012}. Recently, it has been verified that the Shanon entropy of the distribution of white vertical lines, \textit{i.e.}, the recurrence time entropy ($S_{\mathrm{RT}}$), can be used to detect weak chaos in two-dimensional area-preserving maps \cite{SALES10.1063/5.0140613,PhysRevE.109.015202}. It has also been reported that $S_{\mathrm{RT}}$ can detect dynamical transitions on FDS \cite{Gabrick2023}.

The $S_{\mathrm{RT}}$ as a tool for dynamical characterization was originally introduced with no connections to RPs \cite{Little2007} and it provides a good estimate for the Kolmogorov-Sinai entropy \cite{Baptista2010}, for instance. We, on the other hand, consider the RP of a trajectory to estimate the recurrence times and define $S_{\mathrm{RT}}$ as \cite{Little2007, Kraemer2018}
\begin{equation}
    \label{eq:RTE}
    S_{\mathrm{RT}} = -\sum_{v=v_{\textrm{min}}}^{v_{\textrm{max}}} p(v) \ln p(v),
\end{equation}
where $v_{\mathrm{max}}$ and $v_{\mathrm{min}}$ denote the length of the longest and shortest white vertical lines, respectively. The term $p(v) = P(v)/N_{\text{w}}$ represents the relative distribution of white vertical line segments with length $v$, where $N_\text{w}$ is the total number of white vertical line segments and $P(v)$ is the number of white vertical line segments with length $v$ and is given by
\begin{equation}
	\label{eq:wvld}
	P(v) = \sum_{i, j = 1}^{N}R_{i,j - 1}R_{i,j+v}\prod_{k=0}^{v - 1}(1 - R_{i,j+k}).
\end{equation}

For the purposes of this study, we set $v_{\mathrm{min}} = 1$. It is important to carefully evaluate the distribution of white vertical lines [Eq.~\eqref{eq:wvld}], as it might be biased by the border lines, \textit{i.e.}, the lines that begin and end at the border of the RP. The length of these lines might not represent the line's true length due to the finite size of the RP, thus influencing the distribution of white vertical lines and consequently, the $S_{\text{RT}}$ \cite{KRAEMER2019125977}. Therefore, to avoid such border effects, we exclude from the distribution the border lines.

\section{The Riemann-Liouville fractional standard map} \label{sec:mod}

The standard map, also known as the Chirikov-Taylor map or the kicked rotator map, is a two-dimensional, area-preserving map and is a paradigmatic model for investigating the dynamics and essential properties of Hamiltonian systems. Introduced independently by Bryan Taylor \cite{PhysRevLett.35.1306} and Boris Chirikov \cite{CHIRIKOV1979263}, this area-preserving map is described by the following equations:
\begin{equation}
    \label{eq:stdmap}
     \begin{array}{ll}
         p_{n+1} &= p_n - K \sin x_n,  \\
         x_{n+1} &= x_n + p_{n+1}  \mod 2\pi,
     \end{array}
\end{equation}
where $x_n$ and $p_n$ are the canonical position and momentum, respectively, at discrete times $n = 1, 2, \ldots, N$, and $K$ controls the nonlinearity of the map. This map illustrates the Poincaré surface of section for the dynamics of a simple mechanical system known as the kicked rotator. In this system, $x_n$ and $p_n$ represent the angular position and angular momentum, respectively, of the rotator, and $K$ measures the intensity of the periodic kicks applied to the rotator \cite{Lichtenberg2013, Zaslavsky2005}. The differential equation governing the system is given by:
\begin{equation}
\label{eq:difeqstdmap}
	\ddot{x}+K \sin(x) \sum^{\infty}_{n=0} \delta \Bigl(\frac{t}{T}-n \Bigr)=0,
\end{equation}
and the map given by Eq.~\eqref{eq:stdmap} is derived considering the position and momentum just after the $n$th kick.

 Note that in the standard map, the next iteration depends solely on the current state, limiting its ability to model systems with a strong dependence on past states. To overcome this limitation, fractional differential equations (FDEs) \cite{Kilbas2006, Miller1993} have been employed. These equations generalize the conventional differential equation framework by incorporating fractional derivatives, which account for memory effects and non-local interactions \cite{tarasov2024discrete,math9131464,TaraGen2024,BORIN2024114597}. This approach enables the modeling of complex systems with historical dependencies. FDEs have been applied across various scientific fields, including biology \cite{majee2023impact,jan2023dynamical}, electrodynamics \cite{stefanski2020signal,stefanski2019electromagnetic,sikora2018fractional,sikora2018certain,moreles2017mathematical}, and quantum mechanics \cite{atman2020nonlocal,laskin2017time,lim2006fractional,PhysRevE.66.056108}.

 Thus, to obtain a fractional equation of motion for the kicked rotator, we replace the second-order time derivative in Eq. \eqref{eq:difeqstdmap} with the Riemann-Liouville derivative $_0D^{\alpha}_t$ \cite{Kilbas2006, Miller1993}, obtaining the following equation:
\begin{equation}
    \label{eq:fracdifeqstdmap}
    _0D^{\alpha}_t x+K\sin(x) \sum^{\infty}_{n=0} \delta \Bigl(\frac{t}{T}-n \Bigr)=0, 
\end{equation}
where
\begin{equation}
    _0D^\alpha_tx(t) = \frac{1}{\Gamma(2 - \alpha)}\dv[2]{t}\int_0^t\frac{x(\tau)}{(t - \tau)^{\alpha - 1}}\dd\tau.
\end{equation}
By integrating Eq.~\eqref{eq:fracdifeqstdmap} with $\alpha \in (1, 2]$, the Riemann-Liouville fractional standard map (RLFSM) can be written as \cite{EDELMAN20114573,EDELMAN2009279}
\begin{equation}
    \label{eq:RLFSM}
    \begin{aligned}
        p_{n+1} &= p_n - K \sin x_n     \\
        x_{n+1} &= \frac{1}{\Gamma (\alpha )}\sum_{i=0}^{n} p_{i+1}V_{\alpha}(n-i+1) \mod2\pi,      
    \end{aligned}
\end{equation}
where 
\begin{equation}
    V_{\alpha}(m)=m^{\alpha -1}-(m-1)^{\alpha -1}. 
\end{equation}

\begin{figure}[t!]
	\includegraphics[width=\linewidth]{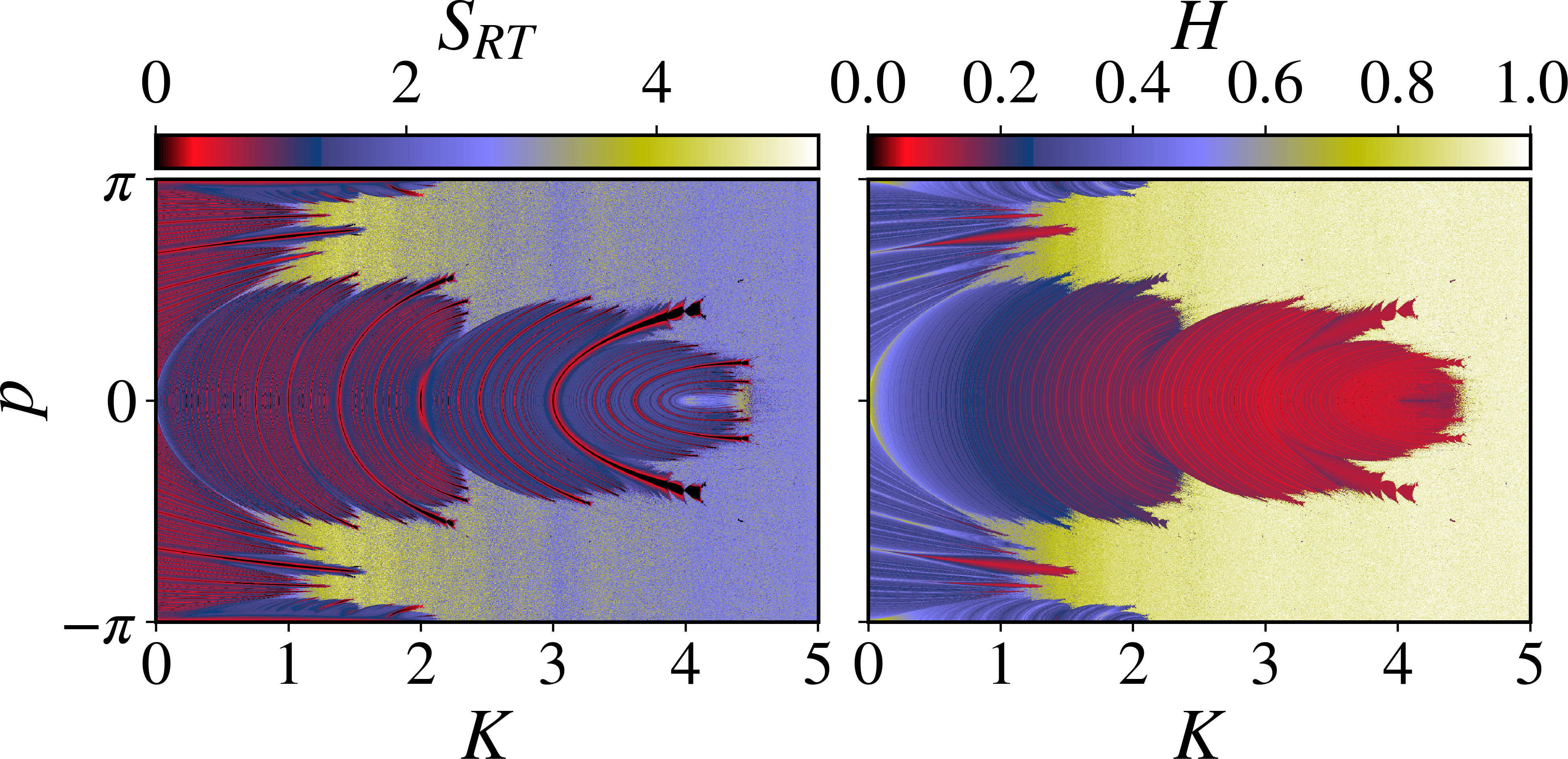}
	\caption{(Left) The recurrence time entropy, $S_{\mathrm{RT}}$, and (right) the Hurst Exponent, $H$, for the RLFSM [Eq.~\eqref{eq:RLFSM}], for a $1000\times1000$ grid of uniformly distributed points in the parameter space $(K, p) \in [0, 5] \times [-\pi, \pi]$, with $x_0 = 0$ and $\alpha = 2$. Each point on the grid was iterated for $N = 1000$ times.}
  \label{fig:pk_alp2}
\end{figure}

\begin{figure*}[t]
	\includegraphics[width=\linewidth]{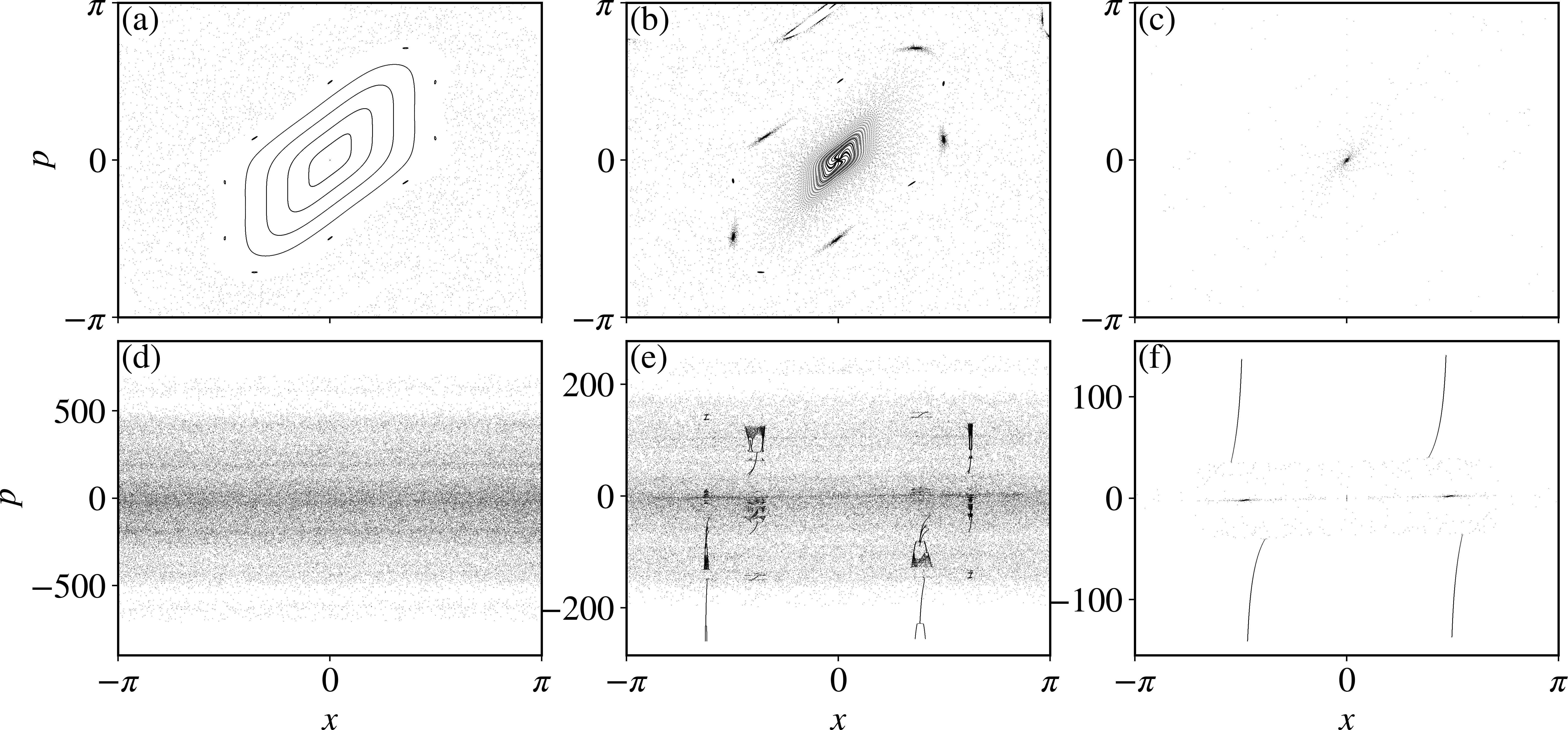}
	\caption{Phase space of the RLFSM for the first $5000$ interaction of $20$ initial conditions, uniformly distributed points along $x=0$ and $p\in [-\pi , \pi]$, with fixed $K=2$ for (a) $\alpha=2$, (b) $\alpha=1.999$, and (c) $\alpha=1.9$; and with fixed $\alpha=1.65$ for (d) $K=7$, (e) $K=4.5$, and (f) $K=4$.}
    \label{fig:phase_space}
\end{figure*}

In the limiting case \( \alpha = 2 \), the RLFSM coincides with the equations for the standard map under the condition \( x_0 = 0 \). Figure \ref{fig:pk_alp2} shows the parameter space of the RLFSM for $K \times p$ on a grid of \( 1000 \times 1000 \) points with \( x_0 = 0 \) and $\alpha = 2$. Each initial condition is iterated \( N = 10^3 \) times, and we calculate the Hurst exponent and recurrence time entropy for each point in the grid. This diagram is known as conservative generalized bifurcation diagram (CGBD) \cite{Manchein2013, Dacosta2022} and is the conservative, \textit{i.e.}, area-preserving, counterpart of traditional bifurcation diagrams of dissipative systems. The CGBD reveals the transitions from regular to chaotic behavior as well as bifurcations as $K$ varies. Figure~\ref{fig:pk_alp2} highlights the similarities between the two observables presented in this paper. Low values of $H$ and $S_{\text{RT}}$ indicate regular (periodic or quasiperiodic) behavior whereas chaotic dynamics is characterized by high values of $H$ and $S_{\text{RT}}$.

To investigate the influence of $\alpha$ on the RLFSM, we perform the phase space analysis of the RLFSM for several values of \( \alpha \) and $K$ (Figure \ref{fig:phase_space}). For $\alpha = 2$ and $K = 2$ [Figure~\ref{fig:phase_space}(a)], the phase space is the typical standard map phase space, naturally. The central stability island is surrounded by smaller islands and all of the regular structures are embedded in the chaotic sea. For values of $\alpha$ close, but different, to 2 [Figures \ref{fig:phase_space}(b) and \ref{fig:phase_space}(c)], the effect of the parameter \( \alpha \) is analogous to a small damping in the standard map \cite{Feudel1996}: the centers of the islands become attracting periodic orbits. However, for smaller values of \( \alpha \), such as \( \alpha = 1.65 \) [Figures~\ref{fig:phase_space}(d)-\ref{fig:phase_space}(f)], we observe completely different behaviors for different $K$. Edelman and coauthors \cite{Edelman2020,10.1063/5.0151812,10.1063/1.4819165, edelman2013new} have demonstrated that the RLFSM can generate attracting asymptotically periodic orbits [Figures \ref{fig:phase_space}(b) and \ref{fig:phase_space}(c)], attracting slow-diverging trajectories, attracting accelerator mode trajectories, and chaotic attractors. They have also shown that the RLFSM exhibits a characteristic type of trajectory, known as cascade of bifurcations type trajectories (CBTTs) [Figure~\ref{fig:phase_space}(e)].

\begin{figure}[t!]
    \includegraphics[width=\linewidth]{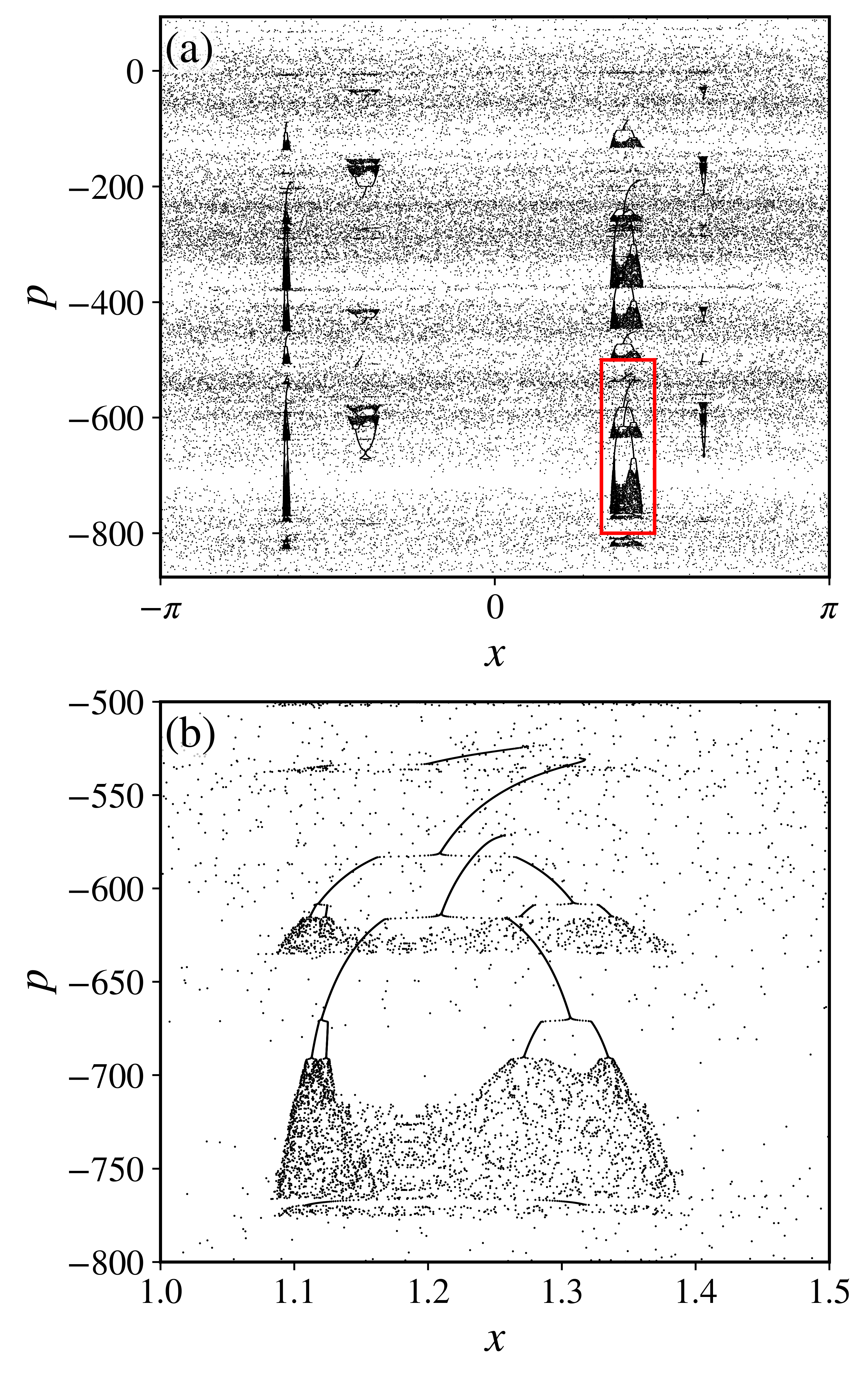}
	\caption{(a) The FSMRL’s phase space for $K = 4.5$ and $\alpha=1.65$ considering $10^5$ iterations on a single trajectory with $(x_0,p_0)=(0,0.3)$. (b) A magnification of one of the CBTT of (a), indicated by the red rectangle.}
        \label{fig:CBTT}
\end{figure}

The CBTT is a characteristic type of regime that, to the best of our knowledge, exists exclusively in fractional dynamical systems. They consist of a sequence of bifurcations in the orbit evolution, which occur not due to variations in system parameters as in conventional dissipative dynamical systems, but rather along a single attracting trajectory during its temporal evolution. In Figure \ref{fig:CBTT}, we observe an example of intermittent CBTT by evaluating the initial condition $(x_0, p_0) = (0.0, 0.3)$ over \( 10^5 \) iterations considering $\alpha=1.65$ and $K=4.5$. This special type of trajectory behaves similarly to a typical chaotic trajectory in a two-dimensional, area-preserving map. Such a trajectory occasionally becomes trapped in a specific region of phase space in which it becomes ``less'' chaotic, \textit{i.e.}, its largest Lyapunov exponent \cite{SZEZECH2005394} and its corresponding Hurst exponent \cite{Borin_Hurst} and recurrence time entropy \cite{SALES10.1063/5.0140613,PhysRevE.109.015202} decreases, for example. Hence the term ``weak chaos'' is used as a reference to the stickiness effect. Therefore, a similar intermittent behavior to the one observed in typical chaotic trajectories in two-dimensional, area-preserving maps is observed in fractional dynamical systems, such as the RLFSM. 
In our case, however, the trappings occur for specific time intervals where the portion of phase space occupied by the orbit is significantly smaller.

In the next section, we aim to use measures used in the characterization of stickiness, such as the recurrence time entropy and the Hurst exponent, to characterize the dynamics of an orbit that follows a CBTT.

\section{CBTT and weak chaos}
\label{sec:results}

A dynamical trap is a region in phase space where an orbit can spend arbitrarily finite long periods behaving not equal but similar to a quasiperiodic orbit even though the overall behavior remains chaotic \cite{ZASLAVSKY2002292}. This leads to the phenomenon of stickiness which is typically characterized using the Lyapunov exponents \cite{PhysRevE.91.062907, Okushima2003_PRL, SZEZECH2005394}. However, in order to calculate the Lyapunov of fractional order systems, it is necessary to extend the definition of the Jacobian matrix to include fractional derivatives and include memory effects in the calculation of the Lyapunov exponents \cite{Wu2015, Li2023}. Recently, several other methods have been proposed to detect sticky orbits such as the use of the entropy of recurrence times (recurrence time entropy) \cite{SALES10.1063/5.0140613, PhysRevE.109.015202} and the Hurst exponent \cite{Borin_Hurst}. Both of these methods relies only on the system's time series, which makes them great tools to study fractional dynamical systems.

A commonly used technique in these methods is finite-time analysis, which provides precise detection of transitions between different dynamical regimes in the orbit. This approach starts by selecting an initial condition \((x_0, p_0)\) and evolving it iteratively to generate a time series of length \(N\). The time series is then divided into windows of size \(T\), where the measure of interest is calculated for each one of these windows.

\begin{figure}[t!]
	\includegraphics[width=\linewidth]{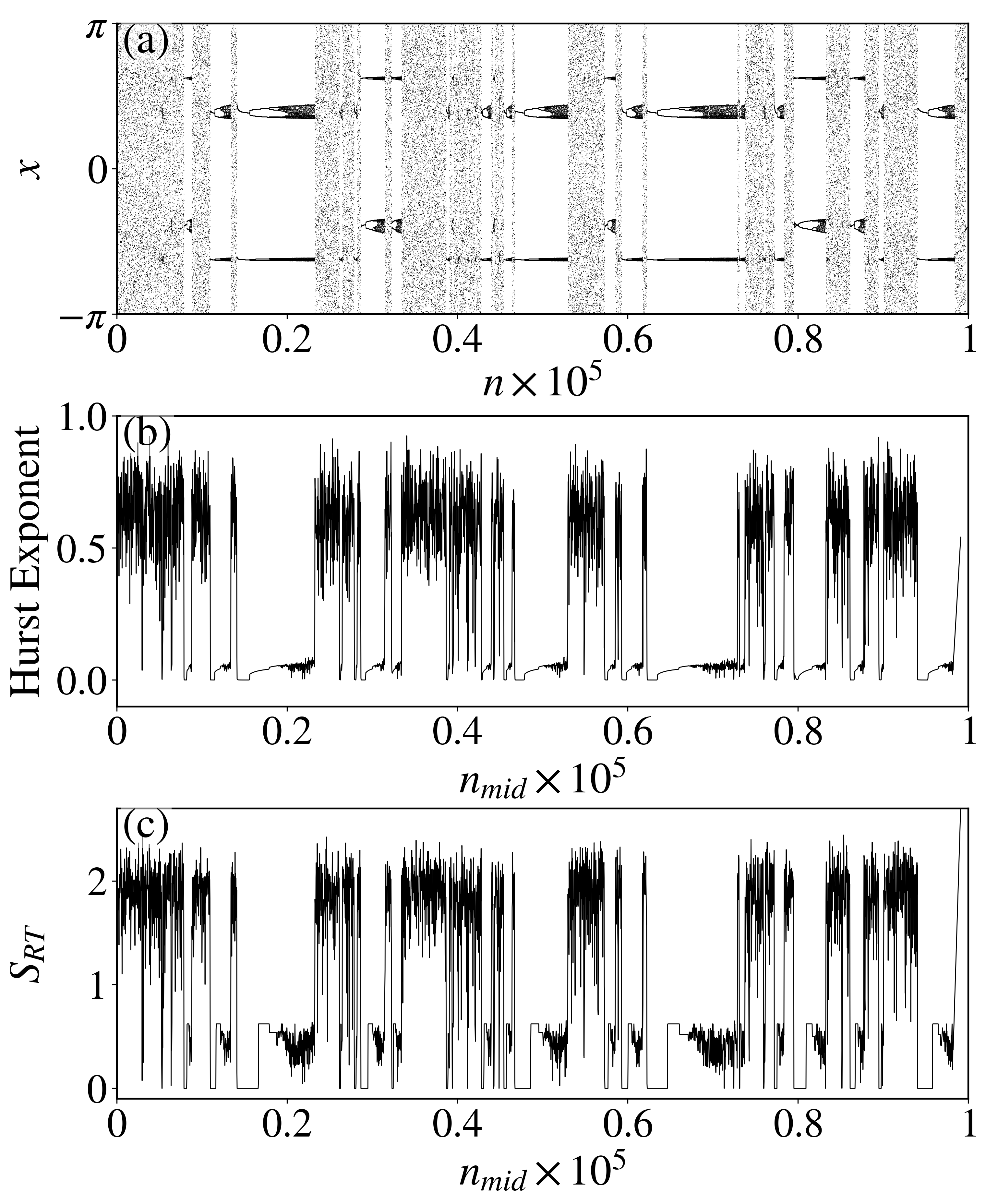}
    \caption{(a) The time series of the variable $x$ for the trajectory shown in Fig.~\ref{fig:CBTT}. (b) The Hurst exponent and (c) the recurrence time entropy as a function of the midpoint of each window.}
        \label{fig:xHRTE}
\end{figure}

We perform such an analysis to quantify CBTTs along a single trajectory of the RLFSM [Eq.~\eqref{eq:RLFSM}] considering the time series of the $x$ variable. We consider an orbit with initial condition \((x_0, p_0) = (0, 0.3)\) and length \(N = 10^5\) [as in Fig.~\ref{fig:CBTT}] and the time series of the $x$ variable is shown in Fig.~\ref{fig:xHRTE}(a). We divide the time series \(X = (x_1, \dots, x_N)\) into \(M = 2^{12}\) partitions. For the \(i\)-th window \((1 < i < N)\), the first and last elements correspond to \(iT\) and \((i+1)T\) of the time series \(X\). The midpoint of each window, given by
\begin{equation}
n_{\text{mid}}^{(i)} = \frac{(i+1)T + iT}{2} = \left(i + \frac{1}{2}\right)T,
\end{equation}
is associated with the respective quantifier which in our case can be either the Hurst exponent or the recurrence time entropy. This analysis is then extended to all partitions. The Hurst exponent and the recurrence time entropy for each partition as a function of the midpoint element \(n_{\text{mid}}\) are shown in Figs.~\ref{fig:xHRTE}(b) and \ref{fig:xHRTE}(c), respectively. The intermittent behavior of an orbit that follows CBTTs is more evident when analyzing the time series of the $x$ variable, for example [Fig.~\ref{fig:xHRTE}(a)]. The orbit abruptly changes its behavior as it evolves in time going from a strong chaotic motion, \textit{i.e.}, the orbit fills the whole $x$ domain, to a seemingly periodic dynamics and to a weaker chaotic motion where the orbit occupies a smaller region, becoming strongly chaotic again. The transition from periodic to weakly chaotic dynamics resembles the period-doubling route to chaos observed in typical dissipative systems, such as the logistic map, for example.

During the chaotic regime, both the Hurst exponent [Fig.~\ref{fig:xHRTE}(b)] and the recurrence time entropy [Fig.~\ref{fig:xHRTE}(c)] have high values. In contrast, as the orbit changes its behavior, both quantifiers exhibit sharp drops to zero, indicating periodic dynamics. After the ``period-doubling'' regime, the orbit reaches the weakly chaotic regime, in which the quantifiers exhibit higher values than the periodic dynamics but smaller than the strong chaotic regime. 
Therefore, both quantifiers detect the intermittent behavior and indicate a weak chaos-like regime that corresponds to the time intervals where the orbit occupies a significantly smaller region in $x$ than it is in the strong chaotic regime.

\begin{figure}[t!]
	\includegraphics[width=\linewidth]{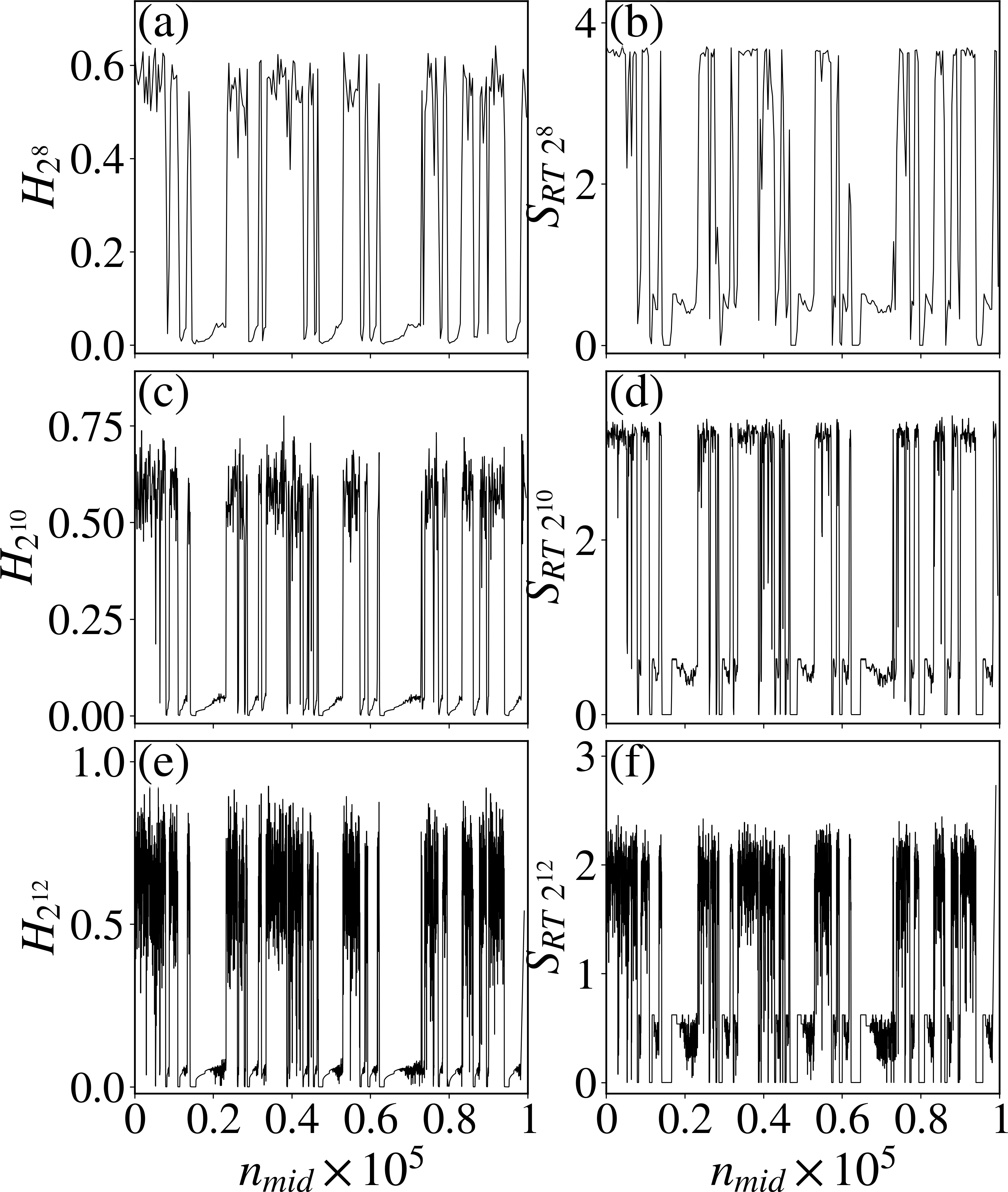}
		\caption{The finite-time Hurst exponent (left column) and the finite-time recurrence time entropy (right column) for time windows of sizes (a) and (b) \(2^8\), (c) and (d) \(2^{10}\), and (e) and (f) \(2^{12}\).}
        \label{fig:HRTE_3winsize}
\end{figure}

Since the value of each quantifier in finite-time analysis depends on the number of divisions of the time series, it is important to understand how different time window sizes influence the results. To address this, Figures \ref{fig:HRTE_3winsize}(a), (c), and (e) show the Finite-Time Hurst Exponent (FTHE) for time windows of sizes \(2^8\), \(2^{10}\), and \(2^{12}\), respectively. Figures \ref{fig:HRTE_3winsize}(b), (d), and (f) present the Finite-Time Recurrence Time Entropy (FTRTE) for the same window sizes as in Figures \ref{fig:HRTE_3winsize}(a), (c), and (e). Our findings indicate that the ability to distinguish between chaotic and CBTT regimes remains consistent across different window sizes. However, as \(M\) increases (resulting in smaller window sizes), the noise level also increases because the quantifiers present a certain level of uncertainty when applied to smaller datasets. Nevertheless, the Hurst exponent and the recurrence time entropy, valuable metrics for evaluating weak chaos in area-preserving dynamical systems, provide a reliable statistical measure of the CBTT effect.

\begin{figure}[t!]
	\includegraphics[width=\linewidth]{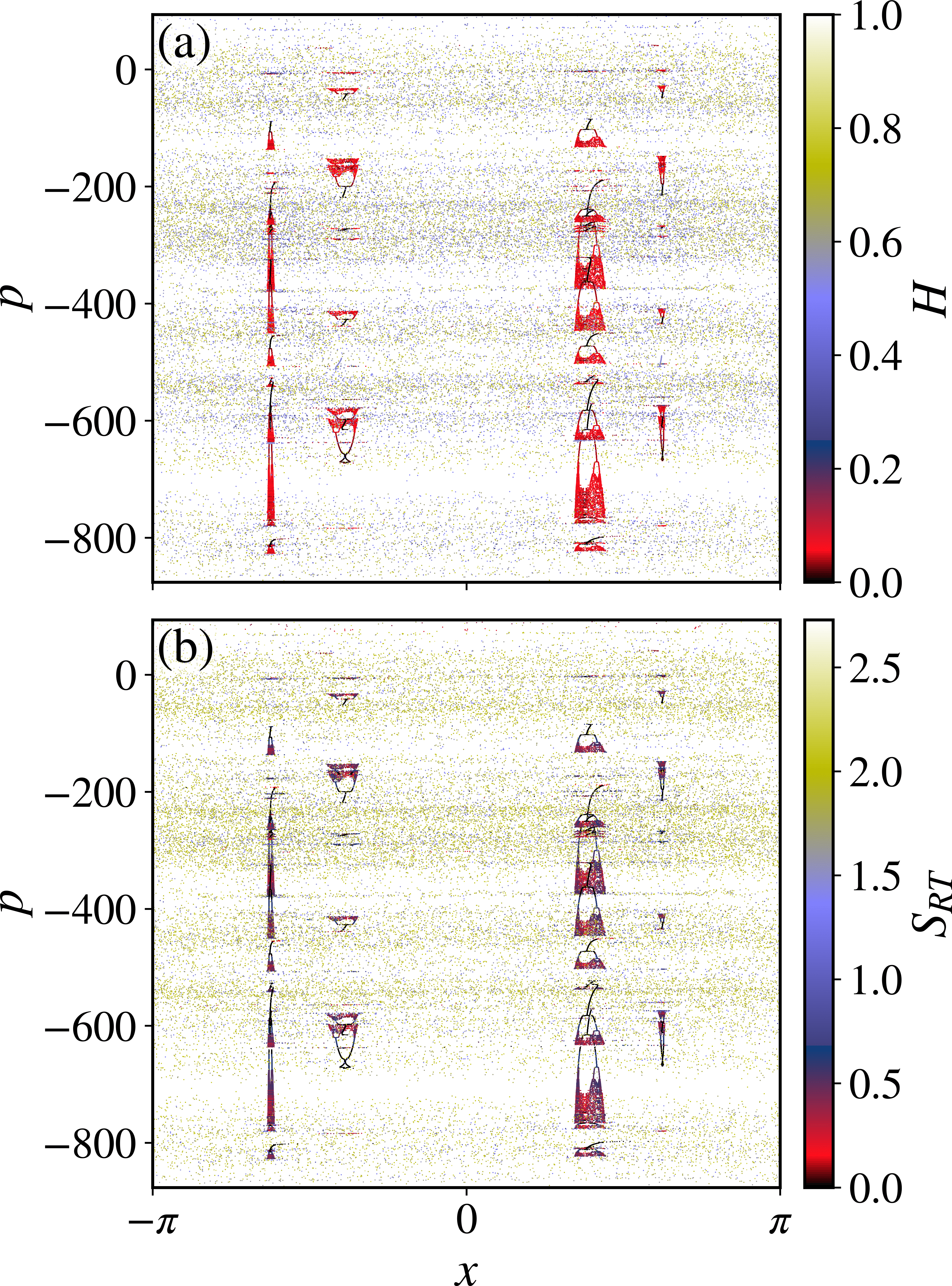}
		\caption{(a) The single trajectory shown in Figure \ref{fig:CBTT} and (b) The Finite-Time Hurst Exponent (FTHE) and (c) Finite-Time Recurrence Time Entropy (FTRTE)  for the same trajectory. Each point is colored according to the quantifier value for the time window in which its coordinate \(x\) is categorized.}
        \label{fig:CBTT_color}
\end{figure}

To identify where the trappings occur in phase space, we perform the previously described finite-time analysis and we plot each point with a color scale according to the quantifier value for the corresponding time window (Fig.~\ref{fig:CBTT_color}). Black color, in our color scale, corresponds to the periodic dynamics and yellow to white corresponds to chaotic dynamics. Weak chaotic dynamics, \textit{i.e.}, the CBTT regime, on the other hand, is indicated by red to purple. During the CBTT regime, the values (and colors) of the quantifiers are clearly different from the large chaotic regions. Therefore, this analysis reveals the cascading effects and provides a detailed and visual characterization of these regions.

\begin{figure}[t!]
	\includegraphics[width=\linewidth]{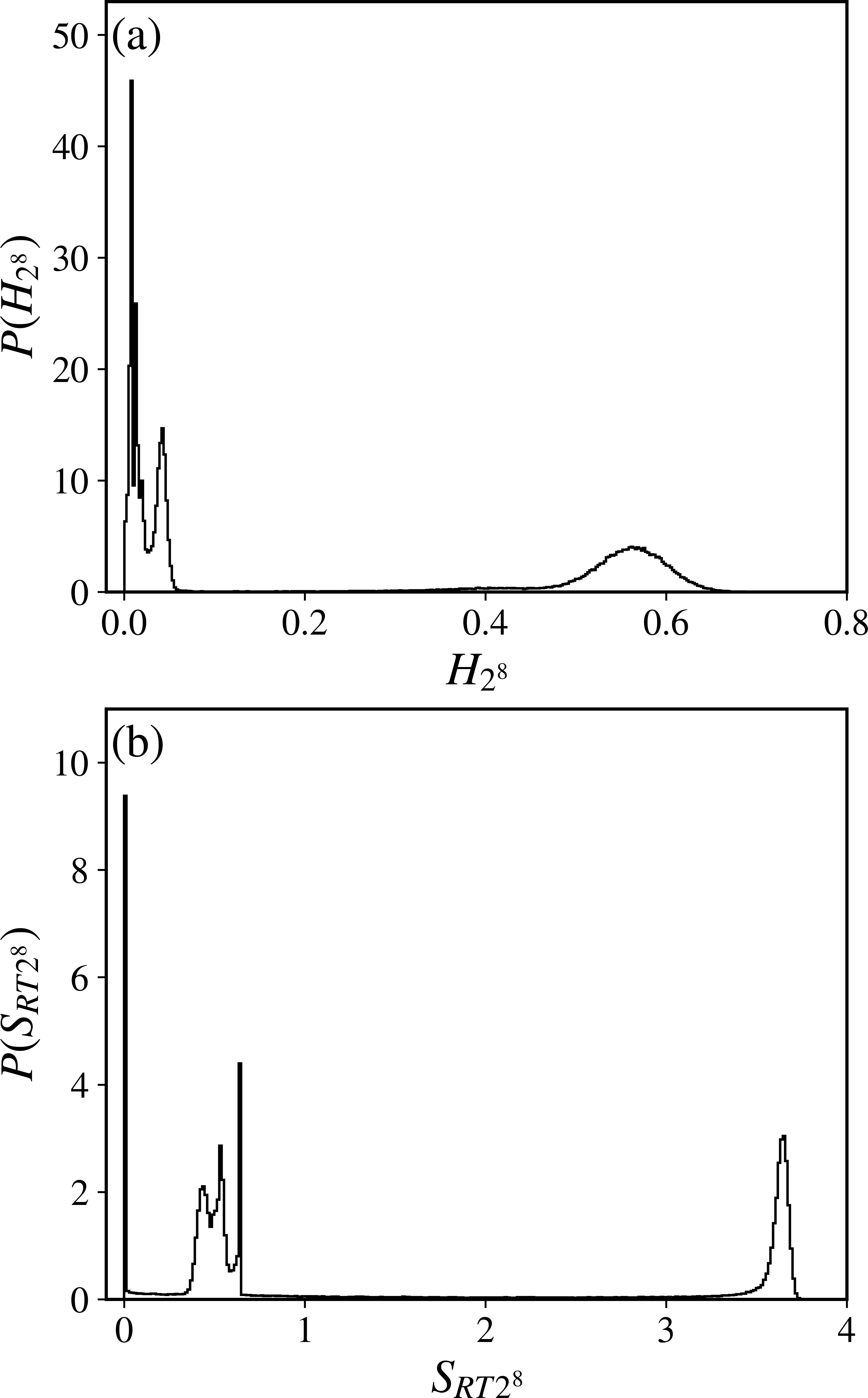}
		\caption{Probability distributions of (a) the finite-time Hurst exponent and (b) the finite-time Recurrence Time Entropy, computed with $M=2^{12}$ partitions on the coordinate $x$ of $10^3$ chaotic orbits of length $N=10^5$, with initial conditions set on the line $x=0$ and $p \in (-\pi, \pi)$ for the FSMRL with $K=4.5$ and $\alpha = 1.65$. Both distributions exhibit multiple peaks, reflecting transitions between chaotic, CBBT, and periodic regimes.}
        \label{fig:distributions}
\end{figure}

In dynamical systems exhibiting the stickiness effect, transitions from fully chaotic motion to various levels within the hierarchical structure of islands-around-islands lead to finite-time distributions of the Hurst exponent and recurrence time entropy \cite{Borin_Hurst, SALES10.1063/5.0140613}, each displaying multiple peaks. Figure~\ref{fig:distributions} shows the probability distributions of these observables for the RLFSM, revealing that fractional dynamics also exhibit multiple peaks. To compute these distributions, we perform a finite-time analysis of \( H \) and \( S_{RT} \) along the evolution of \( 10^3 \) chaotic orbits of length \( N=10^5 \), with initial conditions set on the line \( x=0 \) and \( p \in (-\pi, \pi) \). We considered \( M=2^{12} \) partitions in the \( x \)-coordinate. Using the values of \( H_M \) and \( S_{RT_{M}} \), we construct the probability distributions of the finite-time Hurst exponent, \( P(H_M) \), and the finite-time Recurrence Time Entropy, \( P(S_{RT_{M}}) \), by computing frequency histograms of \( H_M \) and \( S_{RT_{M}} \), respectively.

The Hurst distribution \( P(H_{2^8}) \) [Fig.~\ref{fig:distributions}(a)] exhibits three main peaks. When the orbit is in the chaotic region, the distribution tends to the peak located at higher values of \( H_{2^8} \). In contrast, when the trajectory remains in the CBBT regime, a second and intermediate peak appears at lower values. Finally, when the dynamics reach periodic behavior, the Hurst exponent assumes values corresponding to the highest peak, which appears even closer to zero.
Similarly, the recurrence time entropy distribution \( P(S_{{RT}_{2^8}}) \) [Fig.~\ref{fig:distributions}(b)] follows a similar general pattern to that of the Hurst distribution. However, its primary peak is sharper, and the second peak consists of three smaller sub-peaks, which may indicate different hierarchical levels within the CBBT structure.

\section{Conclusions} \label{sec:conl}

In summary, we have proposed two methods to characterize the dynamics of a fractional system based on its time series, namely, the finite-time Hurst exponents and finite-time recurrence time entropy. Using this methodology, we have shown that due to the similarity between a typical sticky orbit in area-preserving maps and an orbit of a fractional dynamical system, the characterization of CBTTs can be done using the same quantifiers, such as the Hurst exponent and the recurrence time entropy.

Several approaches have been explored in previous studies to detect sticky orbits. However, these methods are not applicable when dealing with fractional maps due to their strong dependence on past states. In this context, the finite-time analysis of the RLFSM using the Hurst exponent and the recurrence time entropy emerges as a powerful alternative for quantifying the CBTT phenomenon since they depend exclusively on the system's time series.

By examining these quantifiers across different time window sizes, we have consistently distinguished between chaotic, periodic, and CBTT regimes, despite the increased noise associated with smaller window sizes. We have shown that chaotic regimes are characterized by higher values of these quantifiers, whereas, for periodic regimes, these quantifiers display small values. The CBTTs regions, on the other hand, display higher values than the periodic regimes but smaller than the chaotic regimes. Therefore, these quantifiers effectively capture the dynamics during the CBTTs regimes, indicating weakly chaotic dynamics during such regimes and enhancing our understanding of the CBTT effect.

The probability distributions of the quantifiers also describe the fractional dynamic. The presence of multiple peaks reflects transitions between chaotic, CBBT, and periodic regimes, providing a characterization of weak chaos in the system. In particular, the structure with three sub-peaks in the intermediate region observed in the recurrence time entropy distribution suggests that recurrence-based measures can distinguish different hierarchical levels within the CBBT regime.


\section*{Funding}

This study was financed, in part, by the São Paulo Research Foundation (FAPESP), Brasil, Process Numbers 2022/03612-6, 2023/08698-9, 2024/06749-8, 2024/09208-8, and 2024/14825-6, and by the National Council for Scientific and Tecnological Development (CNPq), under Grant Nos. 309670/2023-3.

\section*{Author contributions}

The study was conceptualized and designed by DB. Material preparation and data collection were performed by DB and MRS. Data interpretation was done by DB,  JDS, and MRS. The first draft of the manuscript was written by DB. All authors have commented on and suggested previous versions of the manuscript. All authors read and approved the final manuscript

\bibliography{References_FTHE}

\end{document}